\documentclass{article} 
\begin{document} 
\begin{center} 
\title{Quantum mechanical transformation between reference frames - a discursive spacetime diagram approach} 
\author{M. Dance}
\end{center} 

\begin{abstract}
Heisenberg's uncertainty relation means that one observer cannot know an exact position and velocity for another (finite mass) observer.  By contrast, the Poincare transformation of classical special relativity assumes that one observer knows the other's position and velocity exactly.  The present paper describes a simple-minded way to consider the issue using a semiclassical discussion of spacetime diagrams, and draws out some possible implications.  Uncertainties arise in transformations.  A consideration is raised regarding the use of light-cone coordinates.
\end{abstract}

\section{Introduction}

In an early paper on quantum mechanics~\cite{heisenberg}, Heisenberg noted that his uncertainty principle applies to observers as well as to observed systems.  Heisenberg noted that the quantum uncertainties in an observer O1's position and momentum, in another observer O2's coordinates, are subject to his uncertainty principle:

\begin{equation}
\Delta x_{O1}^i   \Delta p_{O1}^i  \geq  \frac{\hbar}{2} .
\end{equation}

This uncertainty relation means that O2 cannot know both O1's position and velocity exactly at the same time in O2's coordinates.  One expects this to be more significant at short time and distance scales, as the uncertainties become a larger proportion of the scales.  By contrast, the Poincare transformation of classical special relativity assumes that one observer knows the other's position and velocity exactly.  

Heisenberg noted that his uncertainty relation was a problem for relativity, and postulated that any practical effect of observer indeterminacy could be eliminated by allowing the observer's mass to approach infinity.  However, a large observer mass warps spacetime.

There have been a number of approaches to the general issue of making special relativity and quantum mechanics consistent, e.g. see~\cite{Kim&Noz06}.  The present paper (based on a 2001 draft) describes a simple way to consider observer transformations using spacetime diagrams, and draws out some possible implications.  Another recent preprint~\cite{WW06} has discussed observer transformations, particularly the notion of a superposition of transformations.  The present paper might possibly be considered as discussing the possible nature of individual relative-velocity components of such a transformation.  The present author also recalls seeing an archived paper around 2001 that discussed the notion of a superposition of reference frames in the context of non-commutative geometry.  A further recent preprint~\cite{Bartlett06} has discussed the concept of quantum reference frames, with particular regard to precision optical experiments with transmitters and receivers at rest relative to each other. Another recent paper~\cite{Rosinger05} has expressed a view that theoretical physics should give more thought to observers and reference frames.

\section{Preliminary discussion}

This paper uses the basic Heisenberg uncertainty relation.  The relation for position and momentum is a fuzzy relation:
\begin{equation}
\Delta x_{O1}^i  \Delta p_{O1}^i   \geq   \frac{\hbar}{2} .
\end{equation}
There is a similar uncertainty relation for time and energy:
\begin{equation}
\Delta E   \Delta t > =  \frac{\hbar}{2} .
\end{equation}

We now drop the label $i$, and consider a spacetime with 1+1 dimensions.  Using the relativistic relation $v = p/\sqrt{p^2 + m_{O1}^2}$ gives in O2's frame (assuming for now that O2 can be said to have a frame, and setting $c=1$ here), we have
\begin{equation} \label{Delta v}
\Delta x_{O1} \Delta v_{O1}   \geq    \frac{m_{O1}^2 \hbar}
					{ 2(p^2 + m_{O1}^2)^{\frac{3}{2}} }
\end{equation}
where $m_{O1}$ is the rest mass of the observer O1.  Close to $c$, the $v$ distribution will be asymmetrical; the higher-$v$ tail will be squashed.  In the non-relativistic limit:
\begin{equation}
\Delta x_{O1} \Delta v_{O1}  \geq  \frac{\hbar}{2m_{O1}}   
\end{equation}

Hypothetically, if O2 could measure O1's velocity with zero uncertainty, $\Delta x_{O1}$ would be infinite.  One might think that this could correspond to some extent with the classical relativistic picture of an infinitely extended observer with clocks and rods extending throughout spacetime, if spacetime has infinite extent.  But by the positions of O1 and O2, we really mean the positions of the origins of those reference frames; O2 has infinite uncertainty about where O1's origin is.  Alternatively, if one can consider a Feynman path integral approach~\cite{feynman&hibbs} to the motion of O1, even if two path endpoints are precisely fixed, there are many possible paths consistent with those endpoints.  A classical limit would pick out the path that extremises the action, but in quantum mechanics one has a superposition of all possible paths, each of which contributes to an interference pattern.

\section{Classical spacetime diagram}

Let O1 and O2 be the origins of classical inertial observers, in a spacetime with one spatial dimension and one time dimension.  We will call O2's coordinates $x_2$ and $t_2$. Suppose that the origin of O1 passes through $x_2 = b$ when $t_2=a$, and that O1's velocity relative to O2 is $v$ in the positive $x_2$ direction.  

Relativity texts show O1's and O2's $t$ and $x$ axes in a simple spacetime diagram.  For technical reasons graphs are not included in this paper, but the description of the spacetime diagram is simple.  O2's axes $t_2$ and $x_2$ are drawn perpendicular to each other.  O1's axes $t_1$ and $x_1$ are lines drawn in the first quadrant of the diagram.   The classical Poincare transformation relating the co-ordinates of O2 to those of O1 is $x_2 = \gamma(x_1 +  vt_1) + b$, $t_2  = \gamma(t_1 + vx_1) + a$,
where $\gamma = 1/\sqrt{1 - v^2}$ and $c=1$.   In O2's frame, O1's time axis is the line 
\begin{equation}
x_2 = v(t_2 - a) + b 
\end{equation}
and O1's spatial axis is the line 
\begin{equation}
x_2 = (t_2 - a)c^2/v + b 
\end{equation}

\section{Spacetime diagram with quantum uncertainty}

What happens to O1's coordinate axes in the simple classical spacetime diagram when O2's measurements of O1 are made subject to the Heisenberg uncertainty principle?  We will assume that O2 has an inertial reference frame and its axes stay the same as above.  We will ignore the backreaction of O2's measurements on O2's reference frame.  The analysis could be considered to be semi-classical.  We will also refer to O1 as meaning O1's origin, where appropriate.

Let us suppose that for times $t_2<0$, O2 has established that O1's velocity in her frame is approximately $v$ in the direction of increasing $x_2$.  Let $a=0=b$.  At $t_2=0$, with uncertainty $\Delta t_2$, O2 observes that O1 passes through $x_2=0$, with an uncertainty $\Delta x_2$.   The $x_2$ measurement introduces an uncertainty of the order of $\Delta v$ into O2's knowledge about O1's velocity, where $\Delta v$ is given by the equality within equation~\ref{Delta v}. We assume that O1's clock reads exactly zero when O2 makes the measurement.  (This assumption could be changed.)

Let us suppose that O2 uses a simple Poincare transformation to estimate O1's coordinates, with the parameters of the transformation now having uncertainties.  In a spacetime diagram, O2 could incorporate the uncertainties by drawing a number of different lines to represent possibilities an O1 time axis, and a number of possible lines to represent an O1 spatial axis.  All of these sets of axes for O1 are options as far as O2 knows.  Because O2 cannot assume or prove O1 to be an inertial observer,  O2 could more appropriately portray O1's axes as smears rather than lines.  But for the sake of simplicity, let us continue with a number of specific sets of possible axes for O1, and understand them to indicate a smearing.

In the new spacetime diagram, O2's axes look perpendicular to each other, just the same as before.  O1's axes are changed.  For any particular value of the parameters (velocity, $a$, and $b$) that one might choose within their uncertainties, a pair of time and space axes for O1 can be drawn.  For example, if we choose a particular value for the velocity within the uncertainty $\Delta v$ from $v$, we could draw with slopes $v$ and $c^2/v$:

-a pair of axes for O1 that intersect at $(-\Delta x_2, -\Delta t_2)$,

-a pair of axes for O1 that intersect at $(-\Delta x_2, +\Delta t_2)$,

-a pair of axes for O1 that intersect at $(+\Delta x_2, -\Delta t_2)$, and

-a pair of axes for O1 that intersect at $(+\Delta x_2, +\Delta t_2)$.

The two most interesting of these intersection points are at $(-\Delta x_2, +\Delta t_2)$ and $(+\Delta x_2, -\Delta t_2)$.  Let us call these points P and Q.

At each point P and Q, a pair of axes for O1 could be drawn for the velocity $v+\Delta v$, and a pair of axes can be drawn for $v-\Delta v$.  But we will focus on the axes for  velocity $v$.  It will be assumed that the speed of light is constant and equal for O2 and O1.  

The time axis for O1 that goes through point Q intersects the spatial axis for O1 that goes through point P.  These two lines roughly define a fuzzy region near O2's (0,0) point, a spacetime region that could be considered as a kind of no man's land, where O2 cannot say much if anything about what O1 perceives as time and what O1 perceives as space.  

The time axis for O1 that goes through point Q - let us call this line L1 - is given by the equation
\begin{equation}
x_2 = v(t_2 + \Delta t_2)   +   \Delta x_2
\end{equation}

The spatial axis for O1 that goes through point P - let us call this line L2 - is given by the equation
\begin{equation}
x_2 = \frac{c^2}{v}(t_2 - \Delta t_2)  -  \Delta x_2
\end{equation}
L1 intersects L2 at $x_2 = x_c$, $t_2 = t_c$, where
\begin{equation} 
\label{xc}
x_c   =   \frac{v( 2\Delta x_2 v/c^2  + 2\Delta t_2  )}
                     {1 - \frac{v^2}{c^2} }
	+ \Delta x_2
\end{equation}
and
\begin{equation}
\label{tc}
t_c   = \frac {   \frac{2\Delta x_2 v}{c^2}  +  
		\Delta t_2 (  1 + \frac{v^2}{c^2} )   }
	        { 1 - \frac{v^2}{c^2} }
\end{equation}

Results for other velocities can be derived by substituting them for $v$ above.  If $(v+\Delta v)$ is substituted in, $x_c$ and $t_c$ contain the products $\Delta x_2 \Delta v$ and $\Delta t_2 \Delta v$ in the numerator, and Heisenberg's uncertainty principle could produce some interesting non-classical results.  

There are two interesting limits: the $v \to c$ limit, and the $v \to 0$ limit.  Let us consider each of these in turn.

\section{The limit as v approaches c}

When $v \to c$, $t_{c}, x_{c}\to \infty$ in equation~\ref{xc}.  A more careful treatment might consider a distribution for relative velocity.  However, noting that for any value of $v$, half of the distribution will be above $v$, and half will be below, it seems intuitively correct that $x_c \to \infty$.  Also, it seems that $\Delta v $ would not significantly affect $x_c$ in the $v \to c$ limit.  In this limit, O2 cannot know much about O1's coordinates in a region covering the two null lines that go through the points P and Q.   The spatial width of O2's region of uncertainty is of the order of $2\Delta x_2$.  And it is expected that $\Delta x_2$ will never be zero, due to a minimum observable length of the order of the Planck length, which has arisen from a generalised uncertainty principle.  Also, as $v \to c$ the "average" spacetime axes for O1, going through O2's origin of coordinates, grow closer to the null line $x_2 = ct_2$.  The extent to which these "average axes" fall within the region defined by P, Q and $(t_c , x_c)$ increases as $v$ increases.

It seems that in a graded way, O1's concept of time and space becomes to some extent less accessible to O2, the faster O1 is moving relative to O2.   By contrast, classical relativity makes a sharp distinction between inertial observers with $v < c$ and entities moving at the speed of light. Classically, any inertial observer with $v < c$ can compute another such observer's coordinates exactly, but this cannot be done for entities moving at the speed of light.

What might an actual value for $x_c$ be, for observers with high relative speed?  If $\Delta t_2$ is zero, $x_{c} \approx \Delta x_2/(1-v)$, which could generate large $x_c$ values.  One might think that this would generate a conflict with experiments; \cite{MM06} has noted that these show Lorentz invariance down to $10^{-18}$~m.  But there need be no conflict.  Nothing in the present paper necessarily influences Lorentz or Poincare invariance; the issue here is rather how to describe transformations between reference systems to begin with.  However, it might be interesting to consider whether quantum uncertainties involved in observers' reference systems might mask Lorentz invariance violations arising from other theories.

It might also be interesting to consider the implications (if any) of these results for the use of light cone coordinates in physics.  If the use of each such coordinate is tantamount to the formal limit of a Poincare transformation multiplied by an overall scale factor, it may be necessary to consider quantum uncertainties in the coordinate transformation; non-commutative geometry probably does this automatically.  This comment is subject to the proviso that the above analysis is a simple semi-classical one, and may not capture all of the relevant effects.  It might however be suggestive.

\section{The zero v limit}

In the nonrelativistic $v \to 0$ limit, we can estimate $x_c$ and $t_c$ by substituting $(v + \Delta v)$ for $v$ in the equations above. We find:

\begin{eqnarray}
t_c (v \to 0)      &\approx &    
		\Delta t_2  +  \frac{2\Delta x_2 \Delta v}{c^2} \nonumber \\
                        &\geq &    \Delta t_2  + \frac{\hbar}{m_{O1}c^2}
\end{eqnarray}
So $t_c$ is nonzero as $v \to 0$, even if $\Delta t_2$ and $\Delta x_2$ are zero; note that they will be at least the relevant Planck scales.

For $x_c$ we have:
\begin{equation}
x_c (v \to 0)      \approx \Delta x_2  +  2\Delta t_2 \Delta v  
		+  \frac{\hbar \Delta v}{m_{O1} c^2}
\end{equation}
Using the Heisenberg uncertainty principle, we have:
\begin{equation}
x_c (v \to 0)     	\geq \Delta x_2  +  \frac{\hbar}{m_{O1} v}
		+  \frac{\hbar \Delta v}{m_{O1} c^2}
\end{equation}
We get a strange term with $v$ in the denominator.  I propose to replace $v$ in the denominator by $\Delta v$.  This can be considered to reflect that fluctuations of order $\Delta v$ will dominate over $v=0$.  Then we have:
\begin{equation}
x_c (v \to 0)	\geq    \Delta x_2  +  \frac{\hbar }{ m_{O1} \Delta v}
		+  \frac{\hbar \Delta v}{m_{O1} c^2} 
\end{equation}
or
\begin{equation} 
x_c (v \to 0)  	\geq   3\Delta x_2  
		+  \frac{1}{2\Delta x_2}\frac{\hbar ^2}{(m_{O1}c)^2}
\end{equation}
The above relations, if correct, imply a nonzero minimum value:
\begin{equation}
x_c (v \to 0)   \geq   \frac{\sqrt{6} \hbar }{m_{O1}c}
\end{equation}

\section{Speculative discussion}

This paper does not attempt to construct a consistent theory of quantum mechanics and special relativity.  Perhaps an observer could be described by a quantum mechanical superposition of classical inertial reference frames; each frame and amplitude in a superposition might be labelled by a pair $(a,p)$, in 1+1 dimensions.  The Poincare transformation might be adaptable accordingly.  A superposition of transformations has been discussed in~\cite{WW06}.  The relevant concepts may depend on how observers are defined.  

The sections above considered a spacetime with one spatial dimension.  It may be interesting to speculate about behaviour in two or more spatial dimensions.  If the simple working above is correct, O2 cannot completely distinguish O1's $t$ and $x$ coordinates.  In 2 spatial dimensions, it seems likely that O2 cannot completely distinguish O1's $t$ and $y$ coordinates.  Perhaps O2 cannot then completely distinguish O1's $x$ and $y$ coordinates, and perhaps a species of non-commutative geometry can reflect this.  It is also to be noted that rotations are part of the Poincare group.  Although non-commutative geometry is generally considered as applicable at high energies, at least one author has previously noted that it may be useful in lower-energy contexts.  

The sections above assumed that the observer O2 could be taken to have a classical inertial reference frame.  However, O2 will in practice be affected by quantum fluctuations.  This would most likely smear out the overall picture even more. O2 will also be subject to a backreaction from its measurement of O1.  These factors have not been considered in the present paper.

The results in this paper may indicate a need to take care when making coordinate transformations.  Rather than simply taking the classical form, some coordinate transformations may inherently add quantum mechanical uncertainties.  In any particular case, the physical content of the transformation may be the determining factor, e.g. whether one is trying to describe what another observer perceives, or not.  On the other hand, it may be that for any given observer, the calculations required of that observer to transform its own coordinates (within its own reference frame/system) might be considered to produce entropy, which in some cases may correspond to the uncertainties discussed above in this paper.  These comments should probably be taken to be limited by what is presently known about the possible interpretations of quantum mechanics.

Relativity could perhaps be modified by considering more realistic observers and configurations of measuring devices than that assumed for classical inertial reference frames.  Such observers might have limited access to information; they might not have passive instantaneous access to all information about events at a distance and their coordinates, as assumed by the classical picture of an observer with access to readings of an infinite number of clocks and rods.  Perhaps a Lagrangian (density) could combine contributions from the observed system, first-tier observing devices, and second-tier observers such as aspects of human observation, in some parametrisation.  A second-tier observer might extrapolate or derive coordinates for spacetime as a whole using information from first-tier devices in a way that uses an explicit model for the second-tier observer's notion of spacetime.  Regarding Lagrangians, it is noteworthy that work is ongoing in relativistic two-body and many-body classical mechanics, e.g.~\cite{Lusanna06}.

\end{document}